\documentclass[]{iopart}
        \usepackage{ amsmathred, latexsym, amssymb, amscd,amsfonts, amsthm,epsfig,color,times,appendix}
\usepackage{graphicx}

\usepackage{pictex}
\usepackage{bbold}
\usepackage[english]{babel}
\newtheorem{theorem}{\bf Theorem}[]

\newtheorem{lem}[]{\bf Lemma}

\begin{document}

\title[]{Statistical analysis of low rank tomography with compressive random measurements}

\author{ \sf \bfseries  Anirudh Acharya, M\u{a}d\u{a}lin Gu\c{t}\u{a}} 
\address{School of Mathematical Sciences, University of Nottingham,
University Park, NG7 2RD Nottingham, United Kingdom}
\date{}

\begin{abstract}
We consider the statistical problem of `compressive' estimation of low rank states with random basis measurements, where the estimation error is expressed terms of two metrics - the Frobenius norm and quantum infidelity. It is known that unlike the case of general full state tomography, low rank states can be identified from a reduced number of observables' expectations. Here we investigate whether for a fixed sample size $N$, the estimation error associated to a `compressive' measurement setup is `close' to that of the setting where a large number of bases are measured.

In terms of the Frobenius norm, we demonstrate that for all states the error attains the optimal rate $rd/N$ with only $O(r \log{d})$ random basis measurements. We provide an illustrative example of a single qubit and demonstrate a concentration in the Frobenius error about its optimal for all qubit states. In terms of the quantum infidelity, we show that such a concentration does not exist uniformly over all states. Specifically, we show that for states that are nearly pure and close to the surface of the Bloch sphere, the mean 
infidelity scales as $1/\sqrt{N}$ but the constant converges to zero as the number of settings is increased. This demonstrates a lack of `compressive' recovery for nearly pure states in this metric. 

\end{abstract}


\section{Introduction}
Recent years have witnessed great experimental progress in the study and control of individual quantum systems \cite{HarocheRaimond,WisemanMilburn}. A common feature of many experiments is the use of quantum state tomography (QST) methods as a 
key tool for validating the results \cite{MIT,tenqubit}.  The aim of QST is to statistically reconstruct an unknown state from the outcomes of repeated measurements performed on identical copies of the state. Among the proposed estimation methods we mention, e.g, variations of maximum likelihood \cite{MLparis,MLrobin,MLmofqubits,HradilML,LocalML}, linear inversion \cite{VogelRisken}, Bayesian inference \cite{BayRDgill,Baykalman}, estimation with incomplete measurements \cite{Maxentropy,Hradilincomplete,maxlikandentropy}, and continuous variables tomography  \cite{LvovskyRaymer}. 

However, for composite systems such as trapped ions, full state tomography becomes  challenging due to the exponential increase in dimension \cite{14qubit}. Therefore, there has been extensive work and significant interest in developing tomography methods for certain lower dimensional families of states. These states are often of experimental interest, and several tomography methods have been developed in this context. For instance, the estimation of low rank states has been considered in the context of compressed sensing (CS) \cite{CSerrorbounds,CSnoRIP,alexandra,alexandragross,CSKalev}, model selection \cite{GutaKypraiosDryden}, and spectral thresholding \cite{rankpenalised,spectralthresholding}. Similarly, the estimation of matrix product states  \cite{Cramer:2010} is particularly relevant for many-body systems, but also for estimating dynamical parameters of open systems \cite{CatanaGutaBouten,GutaKiukas}.

Based on the compressed sensing idea, several recent papers \cite{rankonemeasurements,stablelowrank,AcharyaGuta} consider the problem of estimating low rank states from random measurements. Inspired by the PhaseLift problem, the papers  \cite{rankonemeasurements,phaseliftCandes,phaseliftKeung} consider the case of estimating low rank states from  expectations of rank-one projections sampled randomly from a Gaussian distribution, or a projective t-design, and demonstrate stable compressive recovery with estimation errors of the order of the number of unknown parameters. Compressive quantum process tomography has been considered in this context for unitary 2-designs \cite{KimmelLiu}. In \cite{onblowrank} the analysis is extended to the physically relevant case of random orthonormal basis measurements, and it is shown that a rank-$r$ state can be identified with a large probability for only $O(r \log^3{d})$ such random measurements. Related to this question of low-rank state estimation, work in \cite{KalevBaldwin} conjectures that only a few random bases correspond to \textit{strictly complete POVMs} for low rank states, implying that states of a given low rank can be compressively recovered by measuring a small number of random bases, independent of dimension. 

In this paper we build on the work in \cite{AcharyaGuta}, and consider the statistical problem of estimating low-rank states in the set up of random bases measurements. Instead of choosing a particular estimator, the idea is to investigate the statistical efficiency of an arbitrary optimal estimator, and find whether rank-r states can be estimated from only a few random bases measurements. For this, we consider the behaviour of the mean square error (MSE) with respect to the Frobenius distance between the true state and the estimator $\| \hat{\rho}- \rho\|_2^2$, as well as quantum infidelity $1-F(\hat{\rho},\rho) = 1 - \rm{Tr}\left(\sqrt{\sqrt{\rho}\hat{\rho}\sqrt{\rho}} \right)^2$ in the limit of large number of measurement samples. As discussed below, we show that the answer to the above question is very different for the two distances, emphasising the importance of loss function choice in measurement design.

We first discuss the case of Frobenius distance. According to asymptotic theory \cite{YoungSmith}, in the regime of large number of repetitions the MSE of efficient estimators (e.g. maximum likelihood) $\hat{\rho}$ takes the following expression
\begin{equation} \label{eq.mse.asymptotic}
\mathbb{E} \| \hat{\rho}-\rho \|_2^2 =   \frac{1}{N} {\rm Tr}(I(\rho|\mathcal{S})^{-1} G) + o(N^{-1}).  
\end{equation}
Above, $I(\rho|\mathcal{S})$ is the classical Fisher information associated with the chosen measurement design $\mathcal{S}$ and a local parametrisation of rank-$r$ states, $N$ is the total number of measured systems, 
and $G$ is the positive weight matrix associated with the quadratic approximation of the Frobenius distance in the local parameters. 

The asymptotic MSE \eqref{eq.mse.asymptotic} has been shown to remain robust even with only a few random basis measurements making up the design $\mathcal{S}$ \cite{AcharyaGuta}. This robustness is explained using an argument based on a concentration inequality \cite{winter} for the Fisher information matrix. It is shown in \cite{AcharyaGuta} that certain `least sparse' states of rank-r can be estimated by using only $O(r \log{d})$ settings with only a small increase in the MSE, relative to the setup in which a large number of settings is probed. In this paper the argument using the concentration of the Fisher information is extended to hold for all rank-r states (Theorem \ref{th.concentration}), incorporating the results in \cite{AcharyaGuta}. However, we discuss drawbacks of using a concentration in the Fisher information to derive a corresponding concentration in the MSE. Specifically, for rank-r states that are close to pure with small eigenvalues, we show that such a concentration of the Fisher information does not hold. This difficulty is overcome by proving an \emph{upper bound} on the MSE that holds for all states independently of their spectrum. We show that $\rm{Tr}(I(\rho\vert \mathcal{S})^{-1}G)$ is bounded from above by roughly the number of unknown parameters given that $O(r \log{d})$ random bases constitute the measurement design $\mathcal{S}$. We finally consider the illustrative example of a single qubit state, and analyse the failure of the Fisher concentration for states that are close to pure. We argue that despite a lack of concentration in the Fisher information for such states, the MSE demonstrates the necessary concentration. 

We now discuss the case of infidelity distance. In section \ref{sec:infidelity} we consider the problem of `compressive' state estimation using the quantum infidelity as the error metric. Unlike the Frobenius distance, the quantum infidelity is very sensitive to the misestimation of small eigenvalues. In particular, for states that are close to pure with small eigenvalues, the local expansion of the infidelity in the asymptotic regime is linear in the estimation error of these eigenvalues \cite{MahlerRozema}. Therefore, unlike the Frobenius MSE the expected infidelity is not locally quadratic for all states. We show that for rank-r states with eigenvalues well away from zero, a concentration in the mean infidelity can be demonstrated using a concentration of the Fisher information matrix. While for nearly pure states and random measurements such a concentration of the MSE does not exist. For such states the error scales as $O(1/\sqrt{N})$, and additionally there is no finite number of settings such that the state can be estimated `compressively'.
 
\section{Quantum tomography with random basis measurements}
The setup considered here is motivated by that of multiple ions tomographic (MIT) used in ion-trap experiments \cite{MIT}, with the difference that the measurement bases are \emph{random} rather than Pauli bases. As in MIT, the goal is to statistically reconstruct the joint state of $n$ ions (modelled as two-level systems), from counts data generated by performing a large number of measurements on identically prepared systems. The unknown state $\rho$ is a $d \times d$ density matrix (complex, positive trace-one matrix) where $d = 2^n$ is the dimension of the Hilbert space of $n$ ions. The measurements performed are drawn randomly from the $uniform\ measure$ over orthonormal bases (ONB). Physically, this random setup could be implemented by first rotating the state $\rho$ by a random unitary $U$, after which each atom is measured in the $\sigma_z$ eigenbasis. Let $\mathcal{S} = \{ \mathbf{s}_1, \ldots, \mathbf{s}_k \}$ be the measurement design with $k$ randomly, uniformly distributed measurement bases. Each measurement in a setting $\mathbf{s}$ produces a sequence of $\pm1$ outcomes from each ion $\mathbf{o} = \left( o_1, \ldots ,o_n \right) \in \mathcal{O}_n := \{ +1, -1 \}^n$, whose probability is 
\begin{math} 
p_\rho(\mathbf{o} \vert \mathbf{s} ) := Tr(\rho P_\mathbf{o}^\mathbf{s})
\end{math}, where $P_\mathbf{o}^\mathbf{s}$ is the one-dimensional projection corresponding to the outcome $\mathbf{o}$, in the measurement setting $\mathbf{s}$.

The measurement procedure and statistical model can be summarised as follows. For each setting $\mathbf{s}$ the experimenter performs $m$ repeated measurements and collects the counts of different outcomes $N(\mathbf{o}\vert \mathbf{s})$, so that the total number of quantum samples used is $N := m \times k$. The resulting dataset is a $2^n \times k$ table whose columns are independent and contain all the counts in a given setting.  Our goal is to investigate the statistical efficiency of estimating low rank states in this random measurement setup. We will consider an asymptotic scenario in which the number $m$ of measurement repetitions per setting is large and the mean square error can be characterised in terms of the classical Fisher information, as discussed above.

We assume that the prepared state $\rho$ belongs to the space  of rank $r$ states $\mathbb{S}_r\subset M(\mathbb{C}^d)$, for a fixed rank $r\leq d$. Since the asymptotic mean square error depends only on the local properties of the statistical model, it suffices to consider a parametrisation $\theta\to \rho_\theta$ of a neighbourhood of $\rho$ in  $\mathbb{S}_r$, which can be chosen as follows. In its own eigenbasis $\rho$ is the diagonal matrix of eigenvalues ${\rm Diag}(\lambda_1, \dots, \lambda_r, 0,\dots, 0)$, and any sufficiently close state is uniquely determined by its matrix elements in the first $r$ rows (or columns). Intuitively this can be understood by noting that any rank-r state $\rho^\prime$ in the neighbourhood of $\rho$ can be obtained by perturbing the eigenvalues and performing a small rotation of the eigenbasis; in the first order of approximations these transformation leave the $(d-r)\times (d-r)$ lower-right corner unchanged so 
$$
\rho^{\prime} =
\left(
\begin{array}{ccc}
{\rm Diag}(\lambda_1, \dots, \lambda_r) && 0\\
&&\\
0 &&0
\end{array}
\right)
+ 
 \left(
\begin{array}{ccc}
\Delta_{diag} && \Delta_{off} \\
&&\\
\Delta^{\dagger}_{off}  && O(\|\Delta\|^2 )
\end{array}
\right).
$$
We therefore choose the (local) parametrisation $\rho^\prime = \rho_\theta$ with
\begin{align}
\theta &:= \left( \theta^{(d)}; \theta^{(r)} ;\theta^{(i)} \right) \label{eq.parametrisation} \\
&= ( \rho^\prime_{2,2}, \ldots, \rho^\prime_{r,r}\ ;  {\rm Re}\rho^\prime_{1,2}, \ldots, {\rm Re}\rho^\prime_{r,d}; {\rm Im}\rho^\prime_{1,2}, \ldots, {\rm Im}\rho^\prime_{r,d} ) \in \mathbb{R}^{2rd -r^2-1} \notag
\end{align}
where, in order to enforce a trace-one normalisation, we constrain the first diagonal matrix element to be $\rho^\prime_{1,1}= 1- \sum_{i=2}^{d} \rho^\prime_{i,i}$. After fixing the parametrisation, we now define the statistical model. We denote the set of $k$ uniformly random measurement bases as $\mathcal{S}$. The ions prepared in the unknown state $\rho$ are repeatedly measured $m$ times for each setting in $\mathcal{S}$, so that the overall number of quantum samples $N:=m \times k$. The classical Fisher information associated with a single chosen setting $\mathbf{s}$ is defined as 
\begin{equation}\label{eqn:Fisher}
I(\rho \vert \mathbf{s})_{a,b} :=  \sum_{\mathbf{o}: p(\mathbf{o}\vert \mathbf{s}) >0} \frac{1}{p_{\rho}(\mathbf{o}\vert\mathbf{s})}\frac{\partial p_{\rho}(\mathbf{o}\vert \mathbf{s})}{\partial \theta_{a}} \cdot \frac{\partial p_{\rho}(\mathbf{o}\vert \mathbf{s})}{\partial \theta_{b}}.
\end{equation}

\noindent For a set $\mathcal{S}$ of $k$ settings the Fisher information matrix associated with a single measurement sample from each setting $\mathbf{s}\in \mathcal{S}$ is given by the sum of the individual Fisher matrices $I(\rho \vert \mathbf{s})$, and for later purposes we will denote the average
$
I(\rho \vert \mathcal{S}) = \frac{1}{k} \sum_{\mathbf{s} \in \mathcal{S}} I(\rho \vert \mathbf{s}).
$
The individual matrices can be computed by using definition (\ref{eqn:Fisher}) together with the parametrisation (\ref{eq.parametrisation}).

Since the outcomes from $m$ repeated measurements in a setting $\mathbf{s}$ are i.i.d, when the number of repetitions $m$ is sufficiently large, efficient estimators  of $\theta$ (and hence of $\rho$) from these outcomes have an asymptotically Gaussian 
distribution \cite{YoungSmith}
\begin{equation}\label{eq:normality}
\sqrt{m} (\hat{\theta}-\theta) \approx N(0, I(\rho \vert \mathcal{S})^{-1})
\end{equation}

\noindent where the covariance matrix $I(\rho\vert\mathcal{S})^{-1}$ is the Fisher information associated with a single measurement sample of the set $\mathcal{S}$. We characterise the efficiency of an efficient estimator (e.g. maximum likelihood) in terms of the Frobenius distance. The local expansions of this distance around the state $\rho$ are quadratic in $\theta$, so that 
\begin{equation}\label{eq:localquadratic}
\Vert \rho_{\theta} - \rho_{\theta+\delta \theta} \Vert^2_2 =  (\delta \theta)^{T} G (\delta \theta) + o(\Vert \delta \theta \Vert^2)
\end{equation}

where $G$ is a constant weight matrix for the Frobenius norm. From this and the asymptotic behaviour of efficient estimators, we see that for (reasonably) large $m$, the mean square error scales as   
\begin{equation}\label{eq.mse.fisher}
{\rm MSE}:= \mathbb{E}(\Vert \hat{\rho} - \rho \Vert^2_2) \approx \frac{1}{N} {\rm Tr}( I(\rho \vert \mathcal{S})^{-1}G).
\end{equation}

This trace expression is a measure of the sensitivity of the chosen set of settings $\mathcal{S}$ at $\rho$. We therefore study the behaviour of the MSE, and hence the efficiency of an optimal estimator by considering the behaviour of this quantity with the number of measured settings. We first present a preliminary concentration bound for this quantity ${\rm Tr}( I(\rho \vert \mathcal{S})^{-1}G)$, which extends the results in \cite{AcharyaGuta}. 

\section{Bounds for the MSE}
The bound determines the number of settings $k$ required for the MSE ${\rm Tr}(I(\rho|\mathcal{S})^{-1}G)$ to be concentrated close its optimal value. This result is derived from a concentration of the Fisher information matrix around the mean Fisher information, where the main ingredient is a matrix Chernoff bound for sums of bounded random Hermitian matrices. Since the settings in $\mathcal{S}$ are independent, the Fisher information matrices $I(\rho|{\bf s})$ are independent and this bound is applicable. The Chernoff bound determines how quickly the average information per setting $\frac{1}{k} \sum_{k \in \mathcal{S}} I(\rho \vert \mathbf{s})$ approaches the mean information $\overline{I}$ over all random settings. In terms of the MSE, this translates to determining the number of settings $k$ required for the MSE ${\rm Tr}(I(\rho|\mathcal{S})^{-1}G)$ to be concentrated close the optimal value of ${\rm Tr}(\bar{I}(\rho)^{-1}G)$. We consider states with arbitrary spectrums $\rho := \text{Diag}(\lambda_1,\ldots, \lambda_r,\ldots,0)$, diagonal with respect to its eigenbasis. Due to the unitary symmetry of the random settings design, the eigenbasis can
be chosen to be the standard basis.

\begin{theorem}\label{th.concentration}
Let $\mathcal{S}= \{{\bf s}_1, \dots, {\bf s}_k\} $ be a design with randomly, uniformly distributed measurement bases. 
Let $I_\mathcal{S}:= I(\rho|\mathcal{S})$ be the associated Fisher information, and let $\overline{I}$ be the mean Fisher information over all possible bases, both calculated at the true state $\rho$. For a sufficiently small $\epsilon \geq 0$, the following inequality holds
$$
(1-\epsilon)  {\rm Tr}\left[\overline{I}^{-1}  G \right] \leq {\rm Tr} \left[ I_\mathcal{S}^{-1} G\right] 
\leq (1+\epsilon) {\rm Tr} \left[ \overline{I}^{-1} G \right]  \nonumber
$$
with probability $1-\delta$, provided that the number of measurements performed is $k= \frac{C_1}{\lambda_{\min}(\rho)} \frac{(r+1)}{r} \log(\frac{2D}{\delta})$, 
with $D= 2rd-r^2-1$ the dimension of the space of rank-r states.   
\end{theorem}

The proof of this theorem and further details can be found in the appendix. As mentioned earlier, the main ingredient is a matrix Chernoff bound \cite{winter}, which is used to bound the deviation of $G^{-/2} I(\rho \vert \mathcal{S}) G^{-1/2}$ from the mean $G^{-/2} \bar{I}(\rho) G^{-1/2}$. The number of uniformly random settings $k$ required in the theorem above depends on the following ratio 
\begin{equation} \label{eq:ratio}
\frac{\lambda_{\rm max}}{\lambda_{\rm min}}:= 
\frac{\max_{\mathbf{s}} \lambda_{\rm max}( G^{-1/2}I (\rho \vert \mathbf{s}) G^{-1/2})}{\lambda_{\rm min}( G^{-1/2}\bar{I}(\rho) G^{-1/2})}
\end{equation}
between the largest maximum eigenvalue of $G^{-1/2}I (\rho \vert \mathbf{s}) G^{-1/2}$ over all possible measurements and the minimum eigenvalue of $G^{-1/2}\bar{I}(\rho) G^{-1/2}$. Details of the explicit values of this ratio is left to the appendix. The numerator $\lambda_{\rm max}$ is upper bounded by using the inequality between the quantum and classical Fisher informations \cite{BraunsteinCaves}, as $\lambda_{\rm max}\leq  2/\lambda_{\min}(\rho)$ for $r>1$ and $\lambda_{\rm max}\leq 2$ for $r=1$. While the minimum eigenvalue of $G^{-1/2}\bar{I}(\rho) G^{-1/2}$ is lower bounded using the following lemma. 
\begin{lem}\label{le.average.fisher}
For any rank-r state $\rho$ with an arbitrary spectrum, and the rank-r state $\rho_0$ which has equal non-zero eigenvalues $1/r$ and the same eigenvectors as $\rho$, the following inequality holds between their average Fisher information matrices, evaluated over all possible random measurement settings.
\begin{equation}
\bar{I}(\rho_0) \leq \bar{I}({\rho})
\end{equation}
\end{lem}

The proof is left to the appendix. The matrix $\bar{I}(\rho_0)$ for the equal eigenvalue state has been computed explicitly by using analytic expressions for moments of random unitaries \cite{Collins}, which gives $\lambda_{\rm min}  =\frac{r}{r+1} $ for $r>1$, and $\lambda_{\rm min}=1$ for pure states. Together these give $\frac{\lambda_{\rm max}}{\lambda_{\rm min}} \leq 2\frac{(r+1)}{r}\frac{1}{\lambda_{\min}(\rho)}$ which determines the number of measurement settings in the theorem above. When the state $\rho$ is the equal eigenvalue state $\rho_0$, we get $\lambda_{\min}(\rho_0)=1/r$ and we recover the rate presented in \cite{AcharyaGuta}. 

It was noted in \cite{AcharyaGuta} that deriving a concentration in the MSE via a concentration of Fisher average $I(\rho \vert \mathcal{S})$ provides a pessimistic estimate of the number of settings needed. Simulations in \cite{AcharyaGuta} demonstrated that the MSE concentrates for a much smaller number of settings $k$ than predicted. In the theorem presented above, we note that the dependence of the number of settings on the minimum eigenvalue of $\rho$ suggests a lack of concentration as $\lambda_{\min}(\rho)$ is made arbitrarily small. The number of required settings $k \rightarrow \infty$ in the limit that $\lambda_{\min}(\rho) \rightarrow 0$. This is because the maximum eigenvalue of the Fisher information $I(\rho \vert \mathbf{s})$ over all settings $\mathbf{s}$ becomes arbitrarily large when the rank-r state $\rho$ is arbitrarily close to being pure. However, as we shall demonstrate, this does not reflect the behaviour of the MSE concentration. Instead of deriving a concentration about $\bar{I}(\rho)$ as in the above theorem, we derive a useful upper bound for the MSE that is independent on the spectrum of the state. 

\begin{theorem}\label{th.upper.bound}
Let $\mathcal{S}= \{{\bf s}_1, \dots, {\bf s}_k\} $ be a design with randomly, uniformly distributed measurement bases. Let $I_\mathcal{S}:= I(\rho|\mathcal{S})$ be the associated Fisher information evaluated at $\rho$. For a sufficiently small $\epsilon \geq 0$, the following inequality holds
$$
\text{Tr}[I(\rho \vert \mathcal{S})^{-1}G]\ \leq \ 2 (1+\epsilon) \frac{r+1}{r} D  \nonumber
$$
with probability $1-\delta$, provided that the number of measurements performed is $k= C_2(r+1) \log(2D/\delta)$, 
with $D= 2rd-r^2-1$ the dimension of the space of rank-r states.   
\end{theorem}
 
The upper bound is roughly twice the number of unknown parameters, and although not optimal, it demonstrates that the MSE concentrates below a meaningful threshold given a fixed $O(r \log{D})$ scaling in the number of settings. The key element in the proof of the above theorem is to overcome the potential unboundedness of the maximum eigenvalue of $I(\rho \vert \mathbf{s})$. This is done by bounding $I(\rho \vert \mathbf{s})$ from below over all possible settings $\mathbf{s}$ by matrices whose spectrums are well behaved. This in turn gives us an upper bound for the inverse of the sum $I(\rho \vert \mathcal{S})^{-1}$. 

To this end, we define a new state $\tilde{\rho}$ such that over all possible settings $\mathbf{s}$, we have the following inequality in the Fisher matrices 
 \begin{equation}
 I(\rho \vert \mathbf{s}) \geq \frac{1}{2}I (\tilde{\rho}\vert \mathbf{s}).
 \end{equation}
 
The state $\tilde{\rho}$ is defined to be $\tilde{\rho}:= (\rho + \rho_0)/2$, where $\rho_0$ is the rank-r state with equal $1/r$ eigenvalues, and the same eigenvectors as $\rho$. It is easy to see that $\tilde{\rho}$ has eigenvalues bounded between $(1+1/r)/2$ and $1/2r$, and has the same eigenvectors as $\rho$ by construction. The above inequality then follows from the fact that $\rho \leq 2 \tilde{\rho}$, and from the definition of the Fisher information matrix (\ref{eqn:Fisher}). For any given measurement design $\mathcal{S} = \{ \mathbf{s}_1,\ldots \mathbf{s}_2 \}$, this inequality in the Fisher matrices implies that $I(\rho \vert \mathcal{S}) \geq I(\tilde{\rho} \vert \mathcal{S})/2$. Since the matrix $I(\tilde{\rho} \vert \mathbf{s})$ has eigenvalues that are well behaved over all possible settings $\mathbf{s}$, we can use Theorem \ref{th.concentration} to meaningfully bound the deviation $G^{-1/2}I(\tilde{\rho} \vert \mathcal{S}) G^{-1/2}$ from its mean. In fact, we get that for a sufficiently small $\epsilon \geq 0$, the following inequality holds 
$$
(1-\epsilon)  {\rm Tr}\left[\bar{I}(\tilde{\rho})^{-1}  G \right] \leq {\rm Tr} \left[ I(\tilde{\rho} \vert\mathcal{S})^{-1} G\right] 
\leq (1+\epsilon) {\rm Tr} \left[ \bar{I}(\tilde{\rho})^{-1} G \right]  \nonumber
$$
with probability $1-\delta$, provided that the number of settings $k = C_2(r+1)\log{(2D/\delta)}$. The upper bound in the equation above, combined with the inequality $I(\rho \vert \mathcal{S}) \geq I
(\tilde{\rho} \vert \mathcal{S})/2$  gives us the stated upper bound 
$$
\text{Tr}[I(\rho \vert \mathcal{S})^{-1}G]\ \leq \ 2 (1+\epsilon) \text{Tr}[\bar{I}(\tilde{\rho})^{-1}G] \ \leq \ 2 (1+\epsilon) \frac{r+1}{r} D.
$$ \qed      

Theorem \ref{th.upper.bound} derives a uniform upper bound for all $rank-r$ states irrespective of the eigenvalue spectrum. This demonstrates that sensible bounds exist in the limit of $\lambda_{\min}(\rho) \rightarrow 0$ for a finite number of measurement settings $k$. It is clear that the divergence of maximum eigenvalue $\max_{\mathbf{s}} \lambda_{\max} I(\rho \vert \mathbf{s})$ as $\lambda_{\min}\rightarrow 0$ does not cause a similar divergence in the MSE. Therefore theorem \ref{th.concentration} does not sensibly define a rate for the required number of measured settings $k$ in the limit $\lambda_{\min}\rightarrow 0$. The above theorem does not demonstrate a concentration in the MSE, but only provides a uniform upper bound. However, for the simplified model for a $rank-2$ qubit state, we show that a concentration in the MSE does in fact hold in the limit $\lambda_{\min}(\rho) \rightarrow 0$. 

  \begin{figure}[t]
 \centering
  \includegraphics[scale=0.30]{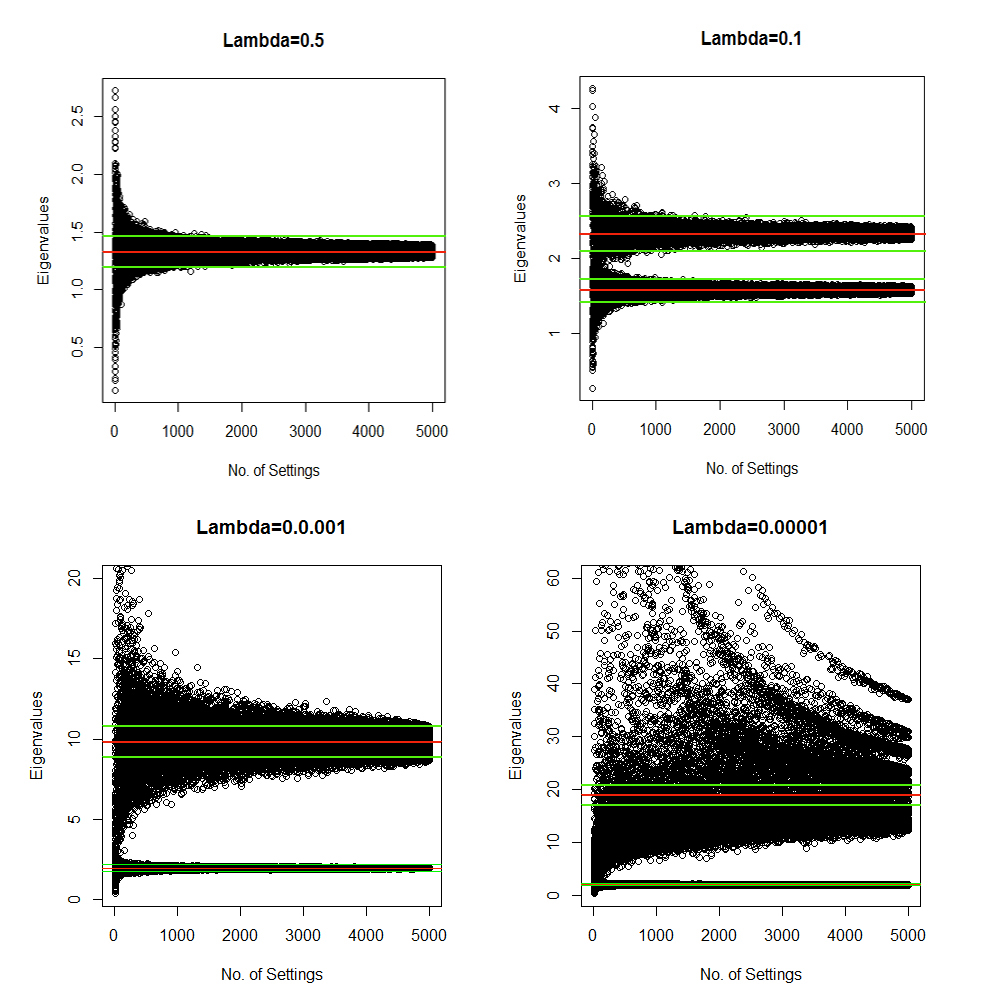}
  \caption{Plots of the eigenvalues of $I({\rho \vert \mathcal{S}})$ for various $k$ random settings. We chose $40$ random single qubit states for each of the four values of $\lambda_2$. The red line indicates the eigenvalues of $\overline{I}(\rho)$, with the green marking the $(1\pm \epsilon) \overline{I}$ deviations ($\epsilon = 0.1$). We observe that as the state becomes purer, the number of settings needed for concentration increases, and in the limit $\lambda_2 \rightarrow 0$ there is a lack of concentration of the largest eigenvalue. }
\label{fig:concentration}
\end{figure}

\section{ The Single Qubit Model}

In this section we work with the simple model of a $rank-2$ qubit state to show that a concentration in the MSE about its optimal holds in the limit $\lambda_{\min} \rightarrow 0$ without requiring the sum $I(\rho \vert \mathcal{S})$ to concentrate about $\overline{I}$. 

\begin{theorem}\label{th.qubit}
Let $\rho$ be a single qubit $rank-2$ state, and let $\mathcal{S} = \{\mathbf{s}_1,\ldots,\mathbf{s}_k\}$ be a uniformly random measurement design. Let $I_{\mathcal{S}} := I(\rho\vert \mathcal{S})$ be the associated Fisher information, and let $\bar{I}(\rho)$ be the mean Fisher information over all possible measurement bases. For any $\epsilon >0$, there exists a finite $k$ such that the following inequality holds for all $\rho$ with high probability

\begin{equation}
\rm{Tr}[I(\rho \vert \mathcal{S})^{-1}G] \leq (1+\epsilon) \rm{Tr}[\bar{I}(\rho)^{-1}G].
\end{equation}
\end{theorem}

In order to investigate the behaviour of the MSE concentration as the spectrum is varied, we consider the generic state $\rho := \lambda_1 \vert 0 \rangle\langle 0 \vert + \lambda_2 \vert 1 \rangle\langle 1 \vert$ diagonal in its eigenbasis. We consider the same local parametrisation as in the previous sections and denote $\theta := \left( \lambda_2, \text{Re}\rho_{1,2}, \text{Im}\rho_{1,2} \right)$. The measurement design consists of random, uniformly distributed measurement bases, and without loss of generality we set the projection vector corresponding to the $+1$ outcome for a given setting $\mathbf{s}$ as:
$$
\vert e^{+1}_{\mathbf{s}} \rangle := \cos{\frac{\theta}{2}} \vert 0 \rangle + e^{i\phi} \sin{\frac{\theta}{2}} \vert1\rangle \ \ \ \ \ \ \ \ 0 \leq \theta \leq \pi \ ,\  0 \leq \phi \leq 2\pi
$$

\noindent The orthogonal vector corresponds to the $-1$ outcome. Therefore, the probabilities $p(\mathbf{o}\vert \mathbf{s})$ corresponding to the two outcomes are $p(+1\vert\mathbf{s}) = (1-\lambda_2) \cos^2{\frac{\theta}{2}} + \lambda_2 \sin^2{\frac{\theta}{2}}$ and $p(-1\vert\mathbf{s}) = (1-\lambda_2) \sin^2{\frac{\theta}{2}} + \lambda_2 \cos^2{\frac{\theta}{2}}$. From equation (\ref{eqn:Fisher}), we evaluate the elements of the Fisher information matrix for a given random measurement setting $\mathbf{s}$. 
\begin{align*}
I(\rho \vert \mathbf{s}) &= 
\begin{pmatrix}
I^{dd}(\rho\vert\mathbf{s}) & I^{dr}(\rho\vert\mathbf{s}) & I^{di}(\rho\vert\mathbf{s}) \\
I^{rd}(\rho\vert\mathbf{s}) & I^{rr}(\rho\vert\mathbf{s}) & I^{ri}(\rho\vert\mathbf{s})  \\
I^{id}(\rho\vert\mathbf{s}) & I^{ir}(\rho\vert\mathbf{s}) & I^{ii}(\rho\vert\mathbf{s})
\end{pmatrix} \\
&= \frac{2}{1-\cos^2(\theta)(1-2\lambda_2)^2}
\begin{pmatrix}
2\cos^2(\theta) & -\cos(\phi)\sin(2\theta) & \sin(\phi)\sin(2\theta) \\
-\cos(\phi)\sin(2\theta) & 2\cos^2(\phi)\sin^2(\theta) & -\sin(2\phi) \sin^2(\theta) \\
\sin(\phi)\sin(2\theta) & -\sin(2\phi) \sin^2(\theta) & 2\sin^2(\phi)\sin^2(\theta) 
\end{pmatrix}
\end{align*} 

 As before  $\mathcal{S}$ is the set of $k$ randomly chosen settings $\mathbf{s}$, and as the settings in $\mathcal{S}$ are independent, the Fisher information matrices $I(\rho \vert \mathbf{s})$ are independent. 
 
  \begin{figure}[t]
 \centering
  \includegraphics[scale=0.30]{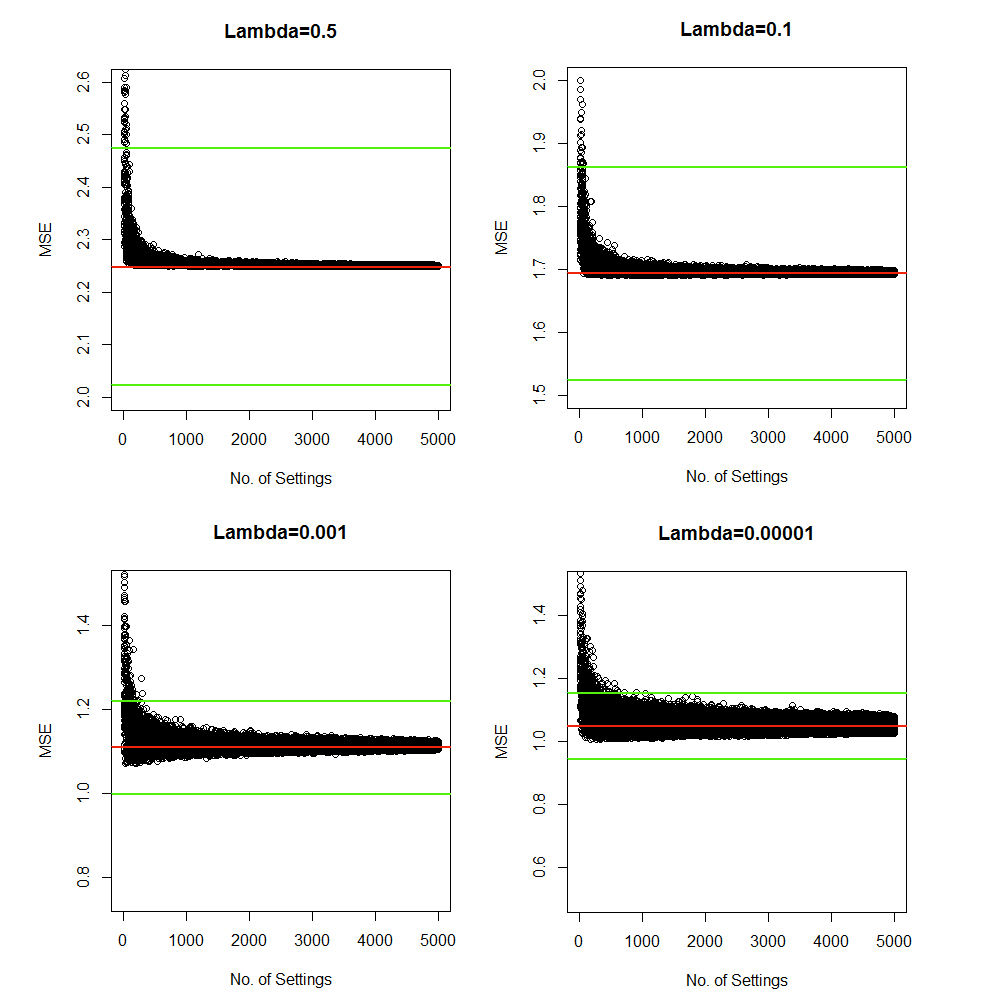}
  \caption{Plots of the MSE of $\text{Tr}I({\rho \vert \mathcal{S}})^{-1}$ for various $k$ random settings. We chose $40$ random single qubit states for each of the four values of $\lambda_2$. The red line indicates the theoretical optimal MSE of $\text{Tr}\bar{I}^{-1}$, with the green marking the $(1\pm \epsilon) \text{Tr}\bar{I}^{-1}$ deviations ($\epsilon = 0.1$). It is easier to observe concentration in the MSE, despite a lack of concentration of $I(\rho \vert \mathcal{S})$ (see Figure \ref{fig:concentration}). Although the number of settings needed for concentration within a prescribed relative error increases with a decrease in $\lambda_2$, there is a limiting value of $k$ as $\lambda_2 \rightarrow 0$ (see text).}
\label{fig:mse.concentration}
\end{figure}

The concentration of the quantity $I(\rho \vert \mathcal{S}) := \frac{1}{k} \sum_{\mathbf{s} \in \mathcal{S}} I(\rho \vert \mathbf{s})$ around the mean Fisher matrix $\overline{I}(\rho)$ is given by Theorem \ref{th.concentration}. We recall that the number of settings $k$ required to bound the deviation from its mean $\bar{I}(\rho)$ depends on the ratio of the eigenvalues
$$
\frac{\lambda_{\text{max}}}{\lambda_{\text{min}}} := \frac{\max_{\mathbf{s}} \lambda_{\max}G^{-1/2}I(\rho \vert \mathbf{s})G^{-1/2}}{\lambda_{\min} G^{-1/2}\overline{I}(\rho)G^{-1/2}}.
$$
\noindent For the simple qubit model these can be explicitly evaluated. For settings with $\theta =0$, the maximum eigenvalue is $\frac{1}{2\lambda_2(1-\lambda_2)}$. This implies that $ \lambda_{\max} \geq \frac{1}{2\lambda_2(1-\lambda_2)}$. This value is a contribution from the $I^{dd}$ element of the Fisher matrix, and tends to infinity as $\lambda_2 \rightarrow 0$. The minimum eigenvalue $\lambda_{\min}$ is a contribution from the $\bar{I}^{rr}$ and the $\bar{I}^{ii}$ term, and tends to a limiting value of 1 when $\lambda_2 \rightarrow 0$. The explicit expressions can be found in Table \ref{table}. Taken together this implies that the ratio becomes unbounded as $\lambda_2 \rightarrow 0$. This is precisely the difficultly characterised in the previous section, and is illustrated in Figure \ref{fig:concentration}, where we plot the eigenvalues of the sum $G^{-1/2}I(\rho\vert \mathcal{S})G^{-1/2}$ for various values of $\lambda_2$ and choices of measurement designs $\mathcal{S}$.  
\bgroup
\begin{table}
\centering
\begin{tabular}{|c| c| c|}
\hline

Fisher Element & Mean & Range \\ [0.5ex]
\hline
$I^{dd}$ & $\dfrac{2\ln{[2(1-\lambda_2)]}-2\ln{[2\lambda_2]}}{(1-2\lambda_2)^3} - \dfrac{4}{(1-2\lambda_2)^2} $& $\left[0, \dfrac{1}{\lambda_2(1-\lambda_2)}\right]$ \\ \hline
$I^{dr}, I^{di}$ &$0$ & $\left[\dfrac{-2}{\sqrt{\lambda_2(1-\lambda_2)}}, \dfrac{2}{\sqrt{\lambda_2(1-\lambda_2)}}\right]$ \\ \hline
$I^{ri}$ & $0$ & $[-2,2]$ \\ \hline
$I^{rr}, I^{ii}$ & $\dfrac{\ln{[2(1-\lambda_2)]}-\ln{[2\lambda_2]}}{(1-2\lambda_2)} -\dfrac{\overline{I}^{dd}}{2}$ & $[0,4]$ \\
\hline
\end{tabular}
\caption{ The mean and range of the elements of the Fisher matrix $I(\rho \vert \mathbf{s})$ as functions of $\lambda_2$. Note that the expressions for the means in the table above are valid for all $\lambda_2 < 0.5$. When $\lambda_2 =0.5$, then all diagonal elements $\bar{I}^{rr/dd/ii}$ have the same value of $4/3$. }   
\label{table}
\end{table}  
\egroup

However we are not interested in the concentration of the Fisher matrix itself, but rather the quantity $\text{Tr}[I(\rho \vert \mathcal{S})^{-1}G]$, and in Figure \ref{fig:mse.concentration} it is seen that the MSE exhibits clear concentration about the optimal. Although the number of settings needed for the MSE to be within $(1 \pm \epsilon)$ of the optimal is seen to increase for smaller values of $\lambda_2$, we shall show that there exists a limiting value of $k$ as $\lambda_2 \rightarrow 0$. To demonstrate this, we consider the concentration of the individual Fisher elements, and directly bound the deviation of $\rm{Tr}[I(\rho\vert \mathcal{S})^{-1}G]$ from its optimal. 

It is clear from Table \ref{table} that the Fisher matrix  elements $I^{rr}, I^{ii}, I^{ri}$ have bounded means and spread even in the limit $\lambda_2 \rightarrow 0$. Their sums can therefore be shown to concentrate around their means using one of several concentration inequalities. For example, we apply Hoeffding's inequality below. 
\newpage

\begin{theorem}\label{th.bounds}
Let $X_1, \ldots, X_k$ be independent random variables such that each $X_i$ is bounded as $a \leq X_i \leq b$, and let $\mu:=\mathbb{E}[X]$. Let $S_k := \frac{1}{k}\sum_i^k X_i $, and $C:=b-a$, then for any $t \geq 0$ and $\tau > 0$ the following inequalities hold, 
\begin{center}
\begin{enumerate}
\item Hoeffding's inequality : $\mathbb{P}(\vert S_k - \mu \vert) \geq t) \leq 2e^{-2kt^2/C^2}$ 
\item Markov's Inequality : $\mathbb{P}(\vert S_k \vert \geq \tau) \leq \frac{\mathbb{E}\vert X\vert}{\tau}$ 
\item Chebyshev's Inequality : $\mathbb{P}(\vert S_k - \mu \vert \geq \tau) \leq \frac{\text{Var}(X)}{\tau^2}$ 
\end{enumerate}
\end{center}
\end{theorem}
From Table \ref{table} we see that $C=4$ for the $I^{rr}, I^{ii}, I^{ri}$ matrix elements. Thus we derive that for any $t \geq 0$, their empirical means are within $\pm t$ of the true value with probability $(1-\delta)$, provided that the number of settings $k \geq (8/t^2) \ln{(2/\delta)}$. Therefore the concentration for these elements is well behaved in the limit $\lambda_2 \rightarrow 0$. While the same inequality can be applied to $I^{rd},I^{id}$ matrix elements when $\lambda_2$ is away from zero, it fails in the limit $\lambda_2 \rightarrow 0$ because their ranges become infinite. However, we make a `\textit{weak law of large numbers}' argument to show that even in this limit, there exists a finite but `sufficiently large' $k$, such that $I^{rd}(\rho\vert \mathcal{S})$ and $I^{id}(\rho\vert \mathcal{S})$ concentrate around their mean. 

The key point is that the random variables $I^{rd},I^{id}$ remain absolutely integrable in limit $\lambda_2 \rightarrow 0$. This is combined with a truncation trick, to show that although the range of these variables in unbounded in the limit, for `sufficiently large' $k$ their empirical means converge in probability to their expected value. We follow the argument presented in \cite{TaoBlog} to demonstrate this. The idea of the truncation method is to split the random variable $I^{rd}$ as
\begin{align*}
I^{rd} &:= I^{rd}_{\leq T} + I^{rd}_{>T} \\
&= I^{rd}\mathbb{1}(\vert I^{rd} \vert {\leq T}) + I^{rd}\mathbb{1}(\vert I^{rd} \vert {> T}), 
\end{align*}
with $T$ being a `truncation parameter' that is chosen appropriately. We shall not be interested in the actual value of $T$, but endeavour only to show that such a method demonstrates the existence of a finite $k$ for which $I^{rd}$ converges in probability to zero. We similarly split the sum 
\begin{align*}
I^{rd}(\rho \vert \mathcal{S}) =
\frac{1}{k}  \sum_{i=1}^k \left[ I^{rd}_{\leq T} (\rho \vert \mathbf{s}_i) + I^{rd}_{> T}(\rho \vert \mathbf{s}_i) \right] =:
I^{rd}_{\leq T}(\rho \vert \mathcal{S}) + I^{rd}_{> T}(\rho \vert \mathcal{S})
\end{align*}
We now bound these two sums using different inequalities. Since the random variable is absolutely integrable even in the limit $\lambda_2 \rightarrow 0$, we can always choose the truncation parameter $T$ such that $\mathbb{E}\vert I^{rd}_{> T} \vert$ is made small, say some $\delta_2 >0$, so that from Markov's inequality (Theorem \ref{th.bounds}) we get 
\begin{equation}\label{eqn.markov}
\mathbb{P}\left(\left\vert  I^{rd}_{> T}(\rho \vert \mathcal{S}) \right\vert \geq \tau \right) \leq \frac{\delta_2}{\tau}. 
\end{equation}
The variable $I^{rd}_{\leq T}$ has bounded spread by construction, and therefore has bounded variance. This allows us to use  Chebyshev's inequality (Theorem \ref{th.bounds}), from which we see that 
\begin{equation}\label{eqn.cheb}     
 \mathbb{P}\left( \left\vert I^{rd}_{\leq T}(\rho \vert \mathcal{S}) \right\vert \geq \tau \right) \leq \frac{\text{Var}(I^{rd}_{\leq T})}{k\tau^2}, 
 \end{equation}
where we use the fact that $\mathbb{E}(I^{rd}_{\leq T}) =0$, since the distribution is symmetric about zero. Clearly $\text{Var}(I^{rd}_{\leq T})$ is bounded, and there exists a finite $k$ such that (\ref{eqn.markov}) and (\ref{eqn.cheb}) together imply
$$
\mathbb{P}\left(\vert I^{rd}(\rho \vert \mathcal{S}) \vert \geq \tau \right) \leq \frac{\delta_2}{\tau} + \frac{1}{\tau^2}.
$$ 
The term $\delta_2$ can be made arbitrarily small by choosing $T$ appropriately, which demonstrates that the sum converges in probability to zero for some finite, but `sufficiently large' $k$. Although the above argument was demonstrated with $I^{rd}(\rho\vert \mathcal{S})$, the same holds for $I^{id}(\rho \vert \mathcal{S})$. This leaves the term $I^{dd}(\rho \vert \mathcal{S})$, which due to the non-integrability, infinite mean and range of $I^{dd}$ in the limit $\lambda_2 \rightarrow0 $, does not concentrate around any finite value. The term $I^{dd}$ contributes the maximum eigenvalue of the Fisher matrix over all settings $\mathbf{s}$, and as mentioned earlier its divergence in the limit $\lambda_2 \rightarrow 0$ is why a concentration inequality of the form of Theorem \ref{th.concentration} does not hold. Collecting the individual bounds for the other matrix elements, we have that for any value of $\lambda_2$ there exists a finite $k$ for which with large probability, the matrix sum $I(\rho \vert \mathcal{S})$ has elements
$$
\begin{pmatrix}
\sum_{i=1}^{k} I^{dd}_i /k & [-\tau, +\tau] & [-\tau, +\tau] \\
 [-\tau, +\tau] &  [\mu- t, \mu+t] & [-t, +t] \\
[-\tau, +\tau] & [-t, +t] & [\mu- t, \mu+t] 
\end{pmatrix}
$$

where $\mu:=\mathbb{E}(I^{rr/ii})$. We can now explicitly evaluate $\text{Tr}[I(\rho \vert \mathcal{S})^{-1}G]$, making the simplifying assumption that $k$ is large enough to ignore terms quadratic in the off-diagonal elements, i.e, in $\tau$ and $t$. Going through the calculation, we get that provided $\sum_{i=1}^{k} I^{dd}_i /k > 1$,
\begin{equation}
\text{Tr}[I(\rho \vert \mathcal{S})^{-1}G] \leq \frac{2k}{\sum_{i=1}^{k} I^{dd}_i} + \frac{4}{\mu - t}.
\end{equation}

In order to show that the MSE is close to optimal as in Theorem \ref{th.qubit}, we require that the term on the right in the above equation is smaller than $(1+\epsilon) \rm{Tr}[\bar{I}(\rho)^{-1}G]$. That is, for some $\epsilon >0$, 
$$
\frac{k}{\sum_{i=1}^{k} I^{dd}_i} + \frac{2}{\mu - t} \leq (1+\epsilon) \left[\frac{1}{\overline{I}^{dd}} + \frac{2}{\mu} \right].
$$

When $\lambda_2$ is sufficiently large, the random variable $I^{dd}$ is bounded and therefore the sum $\sum_{i=1}^{k} I^{dd}_i/k$ concentrates about its mean. In the limit $\lambda_2 \rightarrow 0$ however, $1/\overline{I}^{dd} \rightarrow 0$, which implies that the sum $\sum_{i=1}^{k} I^{dd}_i/k$ does not need to concentrate about its (infinite) mean, but only needs to be larger than a value dependent on $\epsilon$ and $t$. In the limit $\lambda_2 \rightarrow 0$, the Fisher element $I^{dd}$ has a limiting distribution which can be explicitly evaluated. From this distribution and the truncation method it is easy to shown that for any value $C$, there  exists a finite number of settings $k$ such that $\sum I^{dd} /k > C$. This implies that for a given $\epsilon>0$, and for all values of $\lambda_2 \in (0,0.5]$ there always exists a finite number of settings $k$ such that the required concentration holds.

\section{Quantum Infidelity}\label{sec:infidelity}

In this section we consider the problem of 'compressive' state estimation in terms of a different metric, the quantum infidelity
\begin{equation}
1-F(\hat{\rho},\rho) = 1 - \rm{Tr}\left(\sqrt{\sqrt{\rho}\hat{\rho}\sqrt{\rho}} \right)^2.
\end{equation}
 
As briefly hinted at in the introduction, a local expansion of this metric is not locally quadratic uniformly over all states. In particular for states that are well in the interior of the state space the expansion is locally quadratic, while for states with eigenvalues that are close to zero, the infidelity becomes linear \cite{MahlerRozema}. This linear expansion highlights the sensitivity of the infidelity to misestimation of small eigenvalues, and we show that in our setup with uniformly random basis measurements, `compressive' estimation for all states in the sense of Theorem \ref{th.qubit} does not hold for this metric.  To demonstrate this we continue considering the single qubit model from the previous section. We derive a theorem for the concentration of the mean infidelity for states well within the Bloch sphere, and then demonstrate a lack of concentration for nearly pure states. As before, we consider the state $\rho = \rm{Diag}(1-\lambda_2, \lambda_2)$ diagonal in its eigenbasis. For qubits, the infidelity can be expressed as \cite{Geometryofquantumstates}
$$
1-F(\hat{\rho},\rho) = 1-\rm{Tr}(\hat{\rho}\rho) - 2\sqrt{\rm{det}\hat{\rho} \cdot \rm{det}\rho}.
$$
A Taylor expansion of the infidelity about $\rho$ demonstrates that for states within the Bloch sphere (i.e. $\lambda_2$ is well alway from zero), the infidelity is locally quadratic in the (local) parameters
\begin{equation}
1-F(\rho_{\theta},\rho_{\theta+\delta \theta})=(\delta \theta)^{T}G (\delta \theta) + O(\Vert \delta \theta \Vert^3),
\end{equation}    
where $G= \rm{Diag}(1/2\lambda_2(1-\lambda_2),2,2)$ is the corresponding weight matrix. In general for states of arbitrary dimension that have eigenvalues away from zero, the local expansion remains quadratic \cite{MahlerRozema}, and a concentration of expected infidelity is readily established using the techniques in the previous sections. Here we formulate this concentration for the single qubit state considered. Combining the above local expansion with the asymptotic normality of efficient estimators (\ref{eq:normality}), the expected infidelity is given by an expression similar to (\ref{eq.mse.fisher}) 
\begin{equation}\label{eq:quadratic.infidelity}
\mathbb{E}(1-F(\hat{\rho},\rho)) \approx \frac{1}{N}\rm{Tr}(I(\rho \vert \mathcal{S})^{-1}G).
\end{equation}
A concentration of this error term can be demonstrated using the same tools used to establish Theorem \ref{th.concentration}. Concretely, we derive the following theorem.

  \begin{figure}[t]
 \centering
  \includegraphics[scale=0.50]{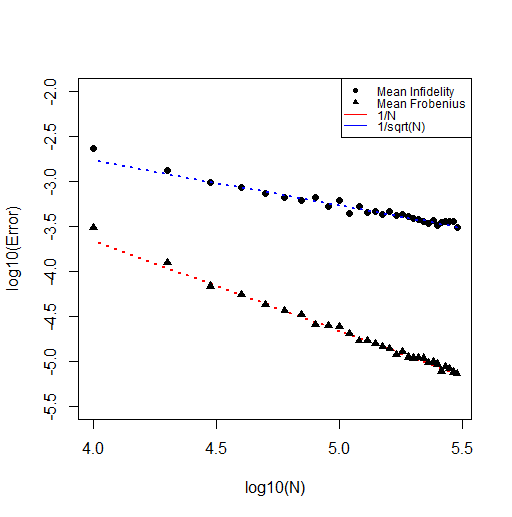}
  \caption{Plots of the mean error in terms of the infidelity and Frobenius norm for the maximum likelihood estimate of a randomly chosen single qubit pure state and random basis measurements. The total number of samples of the state is $N = k \times m$, where $k$ is the number of settings measured and $m=1000$ is the number of repetitions per setting. The number of random basis measured $k$ is varied between 10 and 300. The expected error is approximated over 300 different choices of $k$ randomly chosen settings. The mean Frobenius error demonstrates a $O(1/N)$ scaling, while for the same estimates the Infidelity scales as $O(1/\sqrt{N})$. }
\label{fig:infidelityvsfrobenius}
\end{figure}

\begin{theorem}\label{th.concentration.infidelity}
Let $\mathcal{S}= \{{\bf s}_1, \dots, {\bf s}_k\} $ be a design with randomly, uniformly distributed measurement bases. 
Let $I_\mathcal{S}:= I(\rho|\mathcal{S})$ be the associated Fisher information, and let $\overline{I}$ be the mean Fisher information over all possible bases, both calculated at the single qubit state $\rho$. For a sufficiently small $\epsilon \geq 0$, the following inequality holds
$$
(1-\epsilon)  {\rm Tr}\left[\overline{I}^{-1}  G \right] \leq {\rm Tr} \left[ I_\mathcal{S}^{-1} G\right] 
\leq (1+\epsilon) {\rm Tr} \left[ \overline{I}^{-1} G \right]  \nonumber
$$
with probability $1-\delta$, provided that the number of measurements performed is $k= \frac{C_3}{\rm{det}(\rho)} \log(\frac{2D}{\delta})$, 
with $D= 3$ the dimension of the space of rank-2 qubit states.   
\end{theorem}

Due to the dependence of the number of settings $k$ on the minimum eigenvalue of the true state, the above theorem sensibly demonstrates concentration only when $\lambda_2$ is away from zero. This is similar to the dependence of the number of settings on $\lambda_{\min}(\rho)$ in Theorem \ref{th.concentration}. For the Frobenius norm we demonstrated concentration of the MSE given by $\rm{Tr}(I^{-1}G)$ even in the limit $\lambda_2 \rightarrow 0$. However, in this pure state limit, the local expansion of the infidelity becomes linear in the leading order
\begin{equation}
1-F(\rho_{\theta},\rho_{\theta+\delta \theta}) = |\delta\theta^{d}| + O(\Vert \delta \theta \Vert^2).
\end{equation}
Clearly, for estimates $\hat{\rho}$ in the local neighbourhood of the pure state $\rho= |0\rangle\langle 0|$, the mean error is no longer given by the quadratic expression as in (\ref{eq:quadratic.infidelity}), but is $\mathbb{E}(1-F(\hat{\rho},\rho))= \mathbb{E}(\hat{\theta^d})= \mathbb{E}(\langle 1 \vert \hat{\rho} \vert 1 \rangle)$. Since the dominant error term is linear in the diagonal element of the estimate (in the eigenbasis of the true state), we note that the infidelity is highly sensitive to the misestimation of small eigenvalues \cite{MahlerRozema}. The errors in the estimation of the `rotation parameters' $\theta^r, \theta^i$ however remain quadratic, and therefore exhibit a $O(1/N)$ scaling as in the previous sections. As the interesting contribution to the infidelity is from the estimation errors of the eigenvalue, we consider a simplified single parameter model and assume that only $\theta^d$ is unknown. When the number of repetitions $m$ in a setting $\mathbf{s}$ is sufficiently large, efficient estimators of $\theta^d$ from the outcomes of these measurements have an asymptotically Gaussian distribution 
$\sqrt{m}(\hat{\theta^d} - \theta^d) \approx N(0, \rm{Var}(\hat{\theta^d}))$. Therefore, in this asymptotic limit the expected infidelity $\mathbb{E}(1-F(\hat{\rho},\rho))$ is given by
$$
\mathbb{E}(\hat{\theta^d}) = \frac{1}{\sqrt{2\sigma^2 \pi}}\int_0^\infty \hat{\theta}^d \cdot \rm{exp}\left(-\frac{(\hat{\theta}^d)^2}{2\sigma^2}\right) d\hat{\theta}^d = \sqrt{\frac{2}{\pi}} \sigma,
$$       
where negative estimates of the parameter are set to zero to ensure that $\hat{\rho}$ is a density matrix, and the standard deviation $\sigma =\rm{Var}(\hat{\theta}^d)^{1/2}$. From this asymptotic behaviour of efficient estimators we see that for a large number of repetitions $m$, the mean infidelity scales as 
\begin{equation}\label{eq:local.infidelity.fisher}
\mathbb{E}(1-F(\hat{\rho},\rho)) \approx \sqrt{\frac{2}{\pi N}}\sqrt{I^{dd}(\rho\vert \mathcal{S})^{-1}},
\end{equation}
where the Fisher information $I^{dd}$ corresponding to the diagonal parameter is found in the previous section. From Table \ref{table} and the discussion in the previous section, we know that in the limit $\lambda_2 \rightarrow 0$ the mean Fisher information $\bar{I}^{dd}$ diverges. The Fisher information $I(\rho \vert \mathcal{S})$ for any finite sample of random measurements will therefore not concentrate within $1\pm \epsilon$ of the optimal, implying a lack of concentration in the mean error. In the case of the Frobenius norm, in the limit $\lambda_2 \rightarrow 0$ the dominant error terms contributing to the MSE correspond to the rotation parameters, and this fact ensures a concentration of the MSE even in the pure state limit. While for the infidelity we see that the dominant error terms comes from the estimation of the small eigenvalues, and a concentration of the mean error does not exist in the sense of Theorem \ref{th.qubit}. In general, the local expansion of the infidelity around any rank-$r$ state that is close to pure is linear in the diagonal terms of the estimate \cite{MahlerRozema}. The expected infidelity for such states therefore demonstrates a similar lack of concentration in the corresponding diagonal elements of the Fisher information matrix. 

Furthermore, from (\ref{eq:local.infidelity.fisher}) it is clear that with uniform random measurements the mean infidelity scales as $O(1/\sqrt{N})$ for states that are close to pure. While for states well within the Bloch sphere, Theorem \ref{th.concentration.infidelity} demonstrates a scaling of $O(1/N)$. This poor scaling is observed in Figure \ref{fig:infidelityvsfrobenius}, which plots the expected error in terms of the Infidelity and the Frobenius norm. As discussed in the previous sections, it is seen that the MSE scales as $O(1/N)$ for all states, while it is clear that the mean infidelity demonstrates a $O(1/\sqrt{N})$ scaling for pure states. This scaling has also been demonstrated for the closely related Bures distance error metric. In \cite{KoltchinskiiXia,XiaKoltchinskii}, the minimax Bures error for estimators based on Pauli expectations is shown to scale as $O(1/\sqrt{N})$. This poor scaling along with a lack of concentration is important as many quantum information tasks utilise states that are pure \cite{MahlerRozema}. Several adaptive measurement protocols have been suggested and implemented \cite{MahlerRozema, QiHou,KravtsovStraupe,GranadeFerrie} to improve this scaling. The aim of such adaptive strategies is to make measurements that are close to the eigenbasis of the true state. In our qubit model, for measurements with angle $\theta$ smaller than $O(1/\sqrt{N})$ the Fisher information $I^{dd}(\rho)$ scales as $O(N)$. From (\ref{eq:local.infidelity.fisher}), this gives a $O(1/N)$ scaling of the infidelity even in the limit $\lambda_2 \rightarrow 0$.   

\section{Conclusions}

In this paper we investigated the asymptotic behaviour of the error for an arbitrary optimal estimator in the random measurement setup. Specifically we looked at how the accuracy of efficient estimators depends on the measurement design and the state. We considered two distance measures, the Frobenius norm and the quantum infidelity. In the case of the Frobenius norm, we extended the concentration results in \cite{AcharyaGuta}, and demonstrated that the MSE attains the optimal rate (up to a constant) with only $O(r \log{D})$ random basis measurements for all states of rank $r$. Furthermore, to investigate the behaviour of the MSE concentration for states that are close to pure, we considered the model of a single qubit. We presented an argument to show that concentration in the MSE occurs for all qubit states, despite a lack of concentration in the Fisher information matrix for states close to the surface of the Bloch sphere. 

It remains an open problem if a similar scaling of the MSE exists in the Pauli measurement setup used in standard multiple ions tomography. The application of the tools in the paper to the Pauli setup requires control of the eigenvalues in equation (\ref{eq:ratio}), specifically a lower bound on the minimum eigenvalue $\lambda_{\min}(\bar{I})$. Strong numerical evidence in \cite{AcharyaGuta} suggests that for random measurements the Fisher information may satisfy the required spectral properties. Concentration results for distances other than the Frobenius norm can be in principle derived using similar arguments as long as their local expansions are quadratic in the parameters (see \ref{eq:localquadratic}). However, for the Fidelity (an important measure of error for quantum tomography), it is known that while the scaling is quadratic for states deep in the Bloch sphere, for states close to pure this scaling is linear in the parameters \cite{MahlerRozema}. We demonstrated with a single qubit model that for such nearly pure states and random measurements, the mean infidelity does not concentrate around the optimal for any finite number of settings. This implies that the more random settings, the better, and therefore a lack of `compressive' recovery of the low rank states in this metric.

\appendix
\section*{Appendix}
\section{Proof of Lemma \ref{le.average.fisher}}
\textbf{Lemma.}\textit{
For any rank-r state $\rho$ with an arbitrary spectrum, and the rank-r state $\rho_0$ which has equal non-zero eigenvalues $1/r$ and the same eigenvectors as $\rho$, the following inequality holds between their average Fisher information matrices, evaluated over all possible random measurement settings.
$$
\bar{I}(\rho_0) \leq \bar{I}({\rho})
$$
}
\begin{proof}
For a given random measurement setting $\mathbf{s}$, the probabilities of occurrence of an outcome $\mathbf{o}$ for the two states $\rho_0$ and ${\rho}$ are given by
\begin{align*}
p_0(\mathbf{o}\vert\mathbf{s}) = \sum_{i=1}^{r} \frac{1}{r} \vert \langle e^{\mathbf{o}}_{\mathbf{s}} \vert \lambda_i \rangle \vert^2 \ \ \ ; \ \ 
p_{{\rho}}(\mathbf{o}\vert\mathbf{s}) = \sum_{i=1}^{r} \lambda_i \vert \langle e^{\mathbf{o}}_{\mathbf{s}} \vert \lambda_i \rangle \vert^2 
\end{align*} 
\noindent where $\lambda_i$ and $\vert \lambda_i \rangle$ are the eigenvalues and the eigenvectors of the state $\rho$ respectively.  We now consider states $\rho^{\prime}$, that are constructed by permuting the $r$ non-zero eigenvalues $\lambda_i$ of the state $\rho$, while keeping the eigenvectors fixed. Let $\mathcal{P}$ denote the set of $r!$ such permuted states. The averaged probabilities $p(\mathbf{o}\vert \mathbf{s})$ over all permuted states is given by 
\begin{align*}
\frac{1}{\vert \mathcal{P} \vert} \sum_{\rho^{\prime}\in\mathcal{P}} p_{\rho^{\prime}}(\mathbf{o}\vert\mathbf{s}) &
=\ p_0(\mathbf{o} \vert \mathbf{s}).
\end{align*}
\noindent From the convexity of the function $f(x) = 1/x$ in the interval $(0,+\infty)$, the above equation together with Jensen's inequality implies, 
\begin{equation}\label{eq:jensen}
\frac{1}{p_0(\mathbf{o}\vert \mathbf{s})} \leq \frac{1}{\vert \mathcal{P} \vert} \sum_{\mathcal{\rho^{\prime} \in P} } \frac{1}{p_{\rho^{\prime}}(\mathbf{o} \vert \mathbf{s})} 
\end{equation} 
where we assumed that $p_{\rho^{\prime}} (\mathbf{o}\vert\mathbf{s}) >0$ for all ${\bf o}$. From \ref{eqn:Fisher}, we see that for a setting $\mathbf{s}$, the Fisher matrix in our parametrisation can be written as a sum of $d$ matrices
\begin{equation}
I(\rho^{\prime} \vert \mathbf{s}) = \sum_{\mathbf{o}: p_{\rho^{\prime}} (\mathbf{o}\vert\mathbf{s}) >0} \frac{1}{p_{\rho^{\prime}}(\mathbf{o}\vert \mathbf{s})}  \vert V^{\mathbf{o}}_{\mathbf{s}} \rangle \langle V^{\mathbf{o}}_{\mathbf{s}} \vert
\end{equation}

\noindent where $\vert V^{\mathbf{o}}_{\mathbf{s}} \rangle \in \mathbb{R}^{D}$, with $D= 2rd-r^2-1$, are vectors that depend only on the measurement vectors $\vert e^{\mathbf{o}}_{\mathbf{s}} \rangle$, and the eigenvectors $\vert \lambda_i \rangle$ of the state. Since by construction the eigenvectors for all the states considered above are the same, together with \ref{eq:jensen}, we get for all settings $\mathbf{s}$
\begin{align*}
I (\rho_0| {\bf s}) =\sum_{\mathbf{o}: p_{0} (\mathbf{o}\vert\mathbf{s}) >0} \frac{1}{p_0(\mathbf{o} \vert \mathbf{s})} \vert V^{\mathbf{o}}_{\mathbf{s}} \rangle \langle V^{\mathbf{o}}_{\mathbf{s}} \vert &\leq \frac{1}{\vert \mathcal{P} \vert} \sum_{\mathbf{o}} \sum_{\rho^{\prime} \in \mathcal{P}} \frac{1}{p_{\rho^{\prime}}(\mathbf{o} \vert \mathbf{s})} \vert V^{\mathbf{o}}_{\mathbf{s}} \rangle \langle V^{\mathbf{o}}_{\mathbf{s}} \vert \\
&= \frac{1}{\vert \mathcal{P} \vert} \sum_{\rho^{\prime} \in \mathcal{P}} I(\rho^{\prime} \vert \mathbf{s}).
\end{align*} 
\noindent The inequality holds for settings ${\bf s}$ such that $p_{\rho^{\prime}} (\mathbf{o}\vert\mathbf{s} )>0$ for all ``permuted'' 
states $\rho^\prime$ and all outcomes ${\bf o}$, which holds with probability one under the Haar measure over settings.  Since each $\rho^{\prime}$ is an unitary rotation of the state ${\rho}$, we arrive at the required inequality of the average Fisher matricies by integrating both sides of the above equation over all possible random measurement settings $\mathbf{s}$.  
\end{proof}

\section{Proof of Theorem \ref{th.concentration}}

\begin{proof}[\nopunct]
\noindent The proof of this theorem is similar to the one presented in \cite{AcharyaGuta}. Here we present the important elements of the proof, and refer to \cite{AcharyaGuta} for details. As briefly mentioned in the main text of the paper, the proof of the theorem utilises the following matrix Chernoff bound \cite{winter}, where the random matrices $X_i$ are given by $G^{-1/2}I(\rho \vert \mathbf{s}_i) G^{-1/2}$, with 
${\bf s}_i$ random bases. The proof 

\begin{theorem}{(\textbf{Matrix Chernoff})} Consider a finite sequence $X_1,\ldots,X_k$ of independent, random, positive matrices with dimension $D$, such that $\lambda_{max}(X) \leq R$. For $\mathbb{E}X = M \geq \mu \mathbf{1}$ and $0 \leq \epsilon \leq \frac{1}{2}$, 
\begin{equation}
\mathbb{P} \left\{ \frac{1}{k} \sum_{i=1}^{k} X_i \not\in \biggl[ (1-\epsilon)M, (1+\epsilon)M \biggl] \right\} 
 \leq 2D \cdot \exp{\biggl(-k \cdot \frac{\epsilon^2 \mu}{2R \cdot \log{2}} \biggl)}
\end{equation}
\end{theorem}

\noindent We note that $G^{-1/2}I_{\mathcal{S}}G^{-1/2}$ is a sum of $k$ independent, random, positive matrices. In order to apply the above bound, we need to upper bound the largest eigenvalue of $G^{-1/2}I(\rho \vert \mathbf{s}) G^{-1/2}$ over all measurements. We also need to lower bound the smallest eigenvalue of the expected Fisher information $G^{-1/2}\overline{I}(\rho)G^{-1/2}$. We will first derive these bounds and then derive the result by applying the Chernoff bound. As in the text, we work with the local parametrisation 
\begin{equation}
\theta = \left( \theta^{(d)}, \theta^{(r)}, \theta^{(i)} \right) = \left( \rho_{2,2}, \ldots, \rho_{r,r}; Re \rho_{1,2}, \ldots, Re \rho_{r,d}; Im \rho_{1,2}, \ldots, Im \rho_{r,d} \right) \nonumber
\end{equation}

\noindent where $\rho_{1,1}$ is constrained to enforce the trace-one normalisation. The Fisher information therefore, has the following block structure
$$
I(\rho)=\left(
\begin{array}{ccccc}
I^{dd}(\rho) && I^{dr}(\rho) && I^{di}(\rho) \\ 
&&&&\\
I^{rd}(\rho) && I^{rr}(\rho) && I^{ri}(\rho) \\ 
&&&&\\
I^{id}(\rho) && I^{ir}(\rho) && I^{ii} (\rho)
\end{array}\right)
$$

\noindent with the superscripts identifying the parameters considered; diagonal, real and imaginary. The weight matrix $G$ also has the same block structure with elements  
\begin{equation}\label{eq.G}
G_{a,b}= \rm{Tr}\left[ \frac{\partial \rho_{\theta}}{\partial \theta_{a}} \cdot \frac{\partial \rho_{\theta}}{\partial \theta_{b} } \right]
\end{equation}

In the parametrisation described above, the weight matrix $G$ has the following block diagonal form:
\begin{enumerate}
\item The \textit{diagonal-diagonal} block:
\begin{enumerate}
\item $G^{dd}_{a,b} = 1+ \delta_{a,b}$
\end{enumerate}
\item The \textit{real-real} and \textit{imaginary-imaginary} block:
\begin{enumerate}
\item $G^{rr/ii}_{a,b} = 2 \cdot \delta_{a,b}$
\end{enumerate}
\end{enumerate}

\noindent with the other blocks being zero. We note that both the Fisher, and the weight matrix are of dimension $D:=2rd-r^2-1$. 

\noindent \textit{\textbf{Bound on the smallest eigenvalue}}---As mentioned in the main text, we use Lemma \ref{le.average.fisher} to bound the the smallest eigenvalue from below as
$$
G^{-1/2}\bar{I}(\rho_0)G^{-1/2} \leq G^{-1/2}\bar{I}(\rho)G^{-1/2}.
$$

\noindent Where $\rho_0$ is the state with $r$ equal eigenvalues and the same eigenvectors as the state $\rho$. The explicit form of $\bar{I}(\rho_0)$ is known, and has been evaluated in \cite{AcharyaGuta}, and from it, we see that the minimum eigenvalue is lower bounded by $r/r+1$ for $r>1$ and $1$ for pure states. 

\noindent \textit{\textbf{Bound on the largest eigenvalue}}---We use the inequality $I(\rho \vert \mathbf{s} ) \leq F(\rho)$ between the classical and quantum Fisher informations to bound the largest eigenvalue of $G^{-1/2}I(\rho \vert \mathbf{s} ) G^{-1/2}$ over all measurements by the largest eigenvalue of $G^{-1/2}F(\rho)G^{-1/2}$.  The quantum Fisher information is calculated in the local parameterisation described above and evaluated at the state $\rho = \text{Diag}(\lambda_1,\ldots,\lambda_r,\ldots,0)$, diagonal in its eigenbasis. The details of this calculation can be found in \cite{AcharyaGuta}, and we therefore avoid the repetition and merely state the elements of the matrix. Denoting $r_a, c_a$ to be the row and column positions of the element $a$ of the parameter $\theta$, we have

\begin{enumerate}
\item For the \textbf{\textit{Diagonal-Diagonal}} block with $r>1$,
\begin{enumerate}

\item  $\left. F^{dd}_{a,a} \right|_{\theta}= \frac{1}{\lambda_{r_a}}+\frac{1}{\lambda_1} ~$when $r_a \leq r$
\item $\left. F^{dd}_{a,b} \right|_{\theta = \theta_{0}}= \frac{1}{\lambda_1}$  when $r_a, r_b \leq r$, and $a \neq b$
\end{enumerate}

\item For the \textbf{\textit{Real-Real}} and \textbf{\textit{Imaginary-Imaginary}} blocks:

\begin{enumerate}
\item $\left. F^{rr/ii}_{a,a}\right|_{\theta = \theta_{0}}= \frac{4}{\lambda_{r_a} + \lambda_{c_a}}$ when $r_a < c_a \leq r$ 
\item  $\left. F^{rr/ii}_{a,a}\right|_{\theta = \theta_{0}}= \frac{4}{\lambda_{r_a}}$ when $r_a \leq r, c_a > r$
\end{enumerate}

\end{enumerate}

\noindent The off-diagonal blocks are zero. It is easy to see that the quantum Fisher matrix is upper bounded by the matrix $\frac{1}{\lambda_{\min}(\rho)} G^{dd} \bigoplus \frac{2}{\lambda_{\min}(\rho)} G^{rr} \bigoplus \frac{2}{\lambda_{\min}(\rho)}G^{ii}$. So we can write

$$
G^{-1/2}FG^{-1/2} \leq \frac{1}{\lambda_{\min}(\rho)} \mathbf{1}_{(r-1)} \bigoplus \frac{2}{\lambda_{\min}(\rho)} \mathbf{1}_{(2rd-r^2+r)} \bigoplus \frac{2}{\lambda_{\min}(\rho)} \mathbf{1}_{(2rd-r^2+r)}
$$

\noindent The maximum eigenvalue is therefore upper bounded by $2/\lambda_{\min}(\rho)$ for $r>1$, and $2$ for $r=1$.

\noindent \textit{\textbf{Putting it all together}}-- We can now substitute these values into the matrix Chernoff bound. While the value of the minimum/maximum eigenvalues differ for $r>1$ and $r=1$, we calculate the bound for the case when $r>1$, as this will provide a general bound for the number of settings required that holds even in the case of pure states. Writing $P_{\mathcal{S}}=G^{-1/2}I_{\mathcal{S}}G^{-1/2}$ and $\overline{P} = G^{-1/2}\overline{I}G^{-1/2}$ for notational simplicity, we have for $r>1$
$$
\mathbb{P}\left\{ P_{\mathcal{S}} \not\in \left[ (1- \epsilon) \overline{P}, (1+\epsilon) \overline{P} \right] \right\} 
\leq  2 D \cdot \exp{ \left( -k \frac{r \epsilon^2 \lambda_{\min}(\rho)}{4 (r+1) \cdot \log{2}} \right)} := \delta
$$
 
\noindent Therefore, with probability $1-\delta$ we have that 
$$
(1- \epsilon) \overline{P} \leq P_{\mathcal{S}} \leq (1+\epsilon) \overline{P}
$$

\noindent This can be re-written in the form of inequalities of Mean Square Errors with $\epsilon>0$ sufficiently small
$$
({1- \epsilon}) \rm{Tr} \left( \overline{P}^{-1} \right) \leq \rm{Tr} \left[  P_{\mathcal{S}}^{-1} \right] \leq (1+ \epsilon) \rm{Tr} \left( \overline{P}^{-1} \right)
$$

For a fixed value of $\epsilon$ and $\delta$, we see that the minimum number of settings $k$ required for the above abound to hold with probability greater than $1-\delta$ is 
\begin{equation}
k = \frac{C_1}{\lambda_{\min}(\rho)} \cdot \frac{(r+1)}{r}  \log{\left(\frac{{2D}}{\delta}\right)}  
\end{equation}

\noindent where $C_1 := 4 (\log{2}/\epsilon^2)$ and $D := 2rd-r^2-1$.  

\end{proof}

\section*{References}

\end{document}